\documentclass[preprintnumbers,twocolumn,floats,nofootinbib,superscriptaddress,aps]{revtex4-1}
\usepackage{graphicx,gensymb}
\usepackage{amsmath,amssymb,amsfonts,mathtools}
\usepackage{hyperref}
\usepackage{slashed,wasysym}

\newcommand{\bra}[1]{\langle #1|}
\newcommand{\ket}[1]{|#1\rangle}

\newcommand{\mttwo}{{m_{\rm T2}}}
\newcommand{\ttbar}{{t\bar{t}}}
\newcommand{\mDM}{m_{\rm{DM}}}
\newcommand{\MET}{\slashed{E}_T}

\begin{document}

\setlength{\topmargin}{-1.9cm}
\unitlength=1mm
\textheight=23.75cm

\title{
Extended gamma-ray emission from Coy Dark Matter
}

\preprint{IPPP/13/31, DCPT/13/62, SLAC-PUB-15893}


\author{C\'eline B\oe hm}
\affiliation{Institute for Particle Physics Phenomenology, Durham University, South Road, Durham, DH1 3LE, United Kingdom}
\affiliation{LAPTH, U. de Savoie, CNRS,  BP 110, 74941 Annecy-Le-Vieux, France}
\author{Matthew J.~Dolan}
\affiliation{Theory Group, SLAC National Accelerator Laboratory, Menlo Park, CA 94025, USA\\
{\smallskip \tt  \footnotesize \href{mailto:c.m.boehm@durham.ac.uk}{c.m.boehm@durham.ac.uk}, \href{mailto:mdolan@slac.stanford.edu}{mdolan@slac.stanford.edu}, \href{mailto:christopher.mccabe@durham.ac.uk}{christopher.mccabe@durham.ac.uk}, \href{mailto:michael.spannowsky@durham.ac.uk}{michael.spannowsky@durham.ac.uk}, \href{mailto:c.j.wallace@durham.ac.uk}{c.j.wallace@durham.ac.uk}} \smallskip}
\author{Christopher McCabe}
\author{Michael Spannowsky}
\author{Chris J.~Wallace}
\affiliation{Institute for Particle Physics Phenomenology, Durham University, South Road, Durham, DH1 3LE, United Kingdom}

\begin{abstract}
We show that it is possible for WIMP dark matter to produce a large signal in indirect dark matter searches without producing signals elsewhere. We illustrate our point by fitting the Fermi-LAT extended galactic gamma-ray excess with a simple model of Dirac dark matter that annihilates primarily into $b$ quarks via a pseudoscalar. Current collider constraints are weak while the 14~TeV LHC run will constrain a limited portion of the parameter space. No signal is expected in additional indirect searches or at future direct detection experiments. Our results emphasise the importance of fully understanding potential indirect signals of dark matter as they may provide the only information about the dark matter particle.
\end{abstract}

\maketitle

\section{Introduction}

The precise nature and interactions of particle dark matter remain unknown. Of the many proposed possibilities one particular paradigm has endured: the weakly interacting massive particle (WIMP). WIMPs are assumed to have weak-scale interactions with the Standard Model particles offering the potential for the discovery of dark matter in many channels: direct detection at underground detectors~\cite{Goodman:1984dc}, production at particle colliders~\cite{Cao:2009uw,Beltran:2010ww,Bai:2010hh,Goodman:2010ku} or through indirect searches~\cite{Gunn:1978gr,Stecker:1978du}. Typically, it is assumed that if a signal of WIMP dark matter is found in one of these channels, then a signal will also be found in another channel. Thus the strong limits from the XENON100~\cite{Aprile:2012nq} and LUX~\cite{Akerib:2013tjd} direct detection experiments, which now exclude scattering cross-sections below a typical weak-scale cross-section, have caused some to be pessimistic about the WIMP paradigm.

However, this pessimism is misguided. It is plausible that WIMP dark matter is coy so that it appears at one experiment without producing any other observable signals. We demonstrate this by showing that a simple model of `Coy Dark Matter' (CDM) can explain the recent spatially extended gamma-ray signal of unknown origin from the galactic centre (observed in data from the Fermi-LAT satellite)~\cite{Goodenough:2009gk,Hooper:2010mq,Abazajian:2010zy,Boyarsky:2010dr,Hooper:2011ti,Abazajian:2012pn,Gordon:2013vta,Macias:2013vya}, without producing signals elsewhere. Other examples of CDM include light neutralino dark matter, which can lead to a large signal in the effective number of neutrinos $N_{\textrm{eff}}$ but nowhere else~\cite{Boehm:2012gr,Boehm:2013jpa}. This breakdown of the crossing symmetry relating indirect and direct detection along with collider searches has also been addressed in~\cite{Beltran:2010ww,Goodman:2010ku,Fox:2011pm,LopezHonorez:2012kv,Zhou:2013fla,Profumo:2013hqa}.

Intriguingly, if the extended galactic gamma-ray excess is interpreted in terms of dark matter annihilation, the annihilation cross-section of $\sim10^{-26}\text{ cm}^3\mathrm{s}^{-1}$ required to explain the signal is consistent with that required to obtain the observed relic abundance through thermal freeze-out~\cite{Zeldovich1,Zeldovich2,Chiu1}, a feature of the WIMP paradigm.  Depending on the specifics of the annihilation channel, dark matter with mass between $5$-$50$~GeV provides a good fit to the galactic excess. Previous particle physics oriented studies of this signal have focussed on the $m_{\rm{DM}}\approx10$~GeV region~\cite{Belikov:2010yi,Buckley:2010ve,Zhu:2011dz,Marshall:2011mm,Buckley:2011vs,Boucenna:2011hy,Buckley:2011mm,Hooper:2012cw,Cotta:2013jna,Buckley:2013sca,Hagiwara:2013qya,Fortes:2013ysa,Alves:2013tqa,Modak:2013jya,Arcadi:2013qia}, motivated in part by the persistent signs of a signal in DM direct detection experiments consistent with this mass~\cite{Bernabei:2010mq, Aalseth:2011wp, Angloher:2011uu,Brown:2011dp,Agnese:2013rvf}. 

In this work, we instead consider the higher mass region \mbox{$m_{\rm{DM}}\sim30$}~GeV, which requires that the dominant annihilation is into $b$ quarks. This case is particularly relevant to our discussion since it is for this mass that direct detection experiments are most sensitive. When the dark matter is a Dirac fermion, we show that the observed annihilation cross-section is achieved if the interaction is mediated by a relatively light pseudoscalar with couplings to Standard Model particles that are proportional to the Yukawa couplings (i.e.\ Higgs-like). This coupling structure is well motivated for pseudoscalars from minimal flavour violation (MFV)~\cite{D'Ambrosio:2002ex} and ensures that the dominant annihilation channel is into $b$~quarks. 

Although this scenario produces the observed weak-scale annihilation cross-section, we show that in much of the parameter space, CDM produces no observable signal at other indirect detection, direct detection or collider experiments. With a pseudoscalar mediator, the interaction of dark matter with nucleons is suppressed by the square of the nuclear recoil energy, which is small owing to the non-relativistic nature of the interaction. From a collider perspective, pseudoscalars in this mass range are particularly hard to constrain, since their suppressed couplings to massive vector bosons weaken direct search constraints from LEP and the Tevatron. We find that the greatest sensitivity is afforded by monojet plus missing energy (MET) searches at the LHC, which are sensitive to mediator production followed by decay to dark matter and accompanied by hard QCD radiation.  

Our paper is structured as follows: in Sec.~\ref{sec:lighta} we discuss the extended gamma-ray excess from the galactic centre and find the dark matter mass and annihilation cross-section required to explain it with dark matter annihilation. Following that, we discuss constraints on this scenario from collider searches in Sec.~\ref{sec:LHC} and direct and other indirect detection searches in Sec.~\ref{sec:directdetection}.

\section{The extended gamma-ray excess}
\label{sec:lighta}

Owing to the large dark matter number density there, one of the most promising places to look for dark matter annihilation products is a small $(\sim0.1\text{ kpc})$ region centred on the galactic centre. Evidence for a spatially extended excess of gamma-rays in this region was initially found in~\cite{Goodenough:2009gk} and subsequently confirmed by several independent analyses~\cite{Hooper:2010mq,Boyarsky:2010dr,Hooper:2011ti,Abazajian:2010zy,Abazajian:2012pn,Gordon:2013vta,Macias:2013vya}. A spectrally and morphologically similar excess has also been reported at more extended distances from the galactic plane~\cite{Hooper:2013rwa,Huang:2013pda}. 

\begin{figure}[t!]
\centering
\includegraphics[height=0.84\columnwidth]{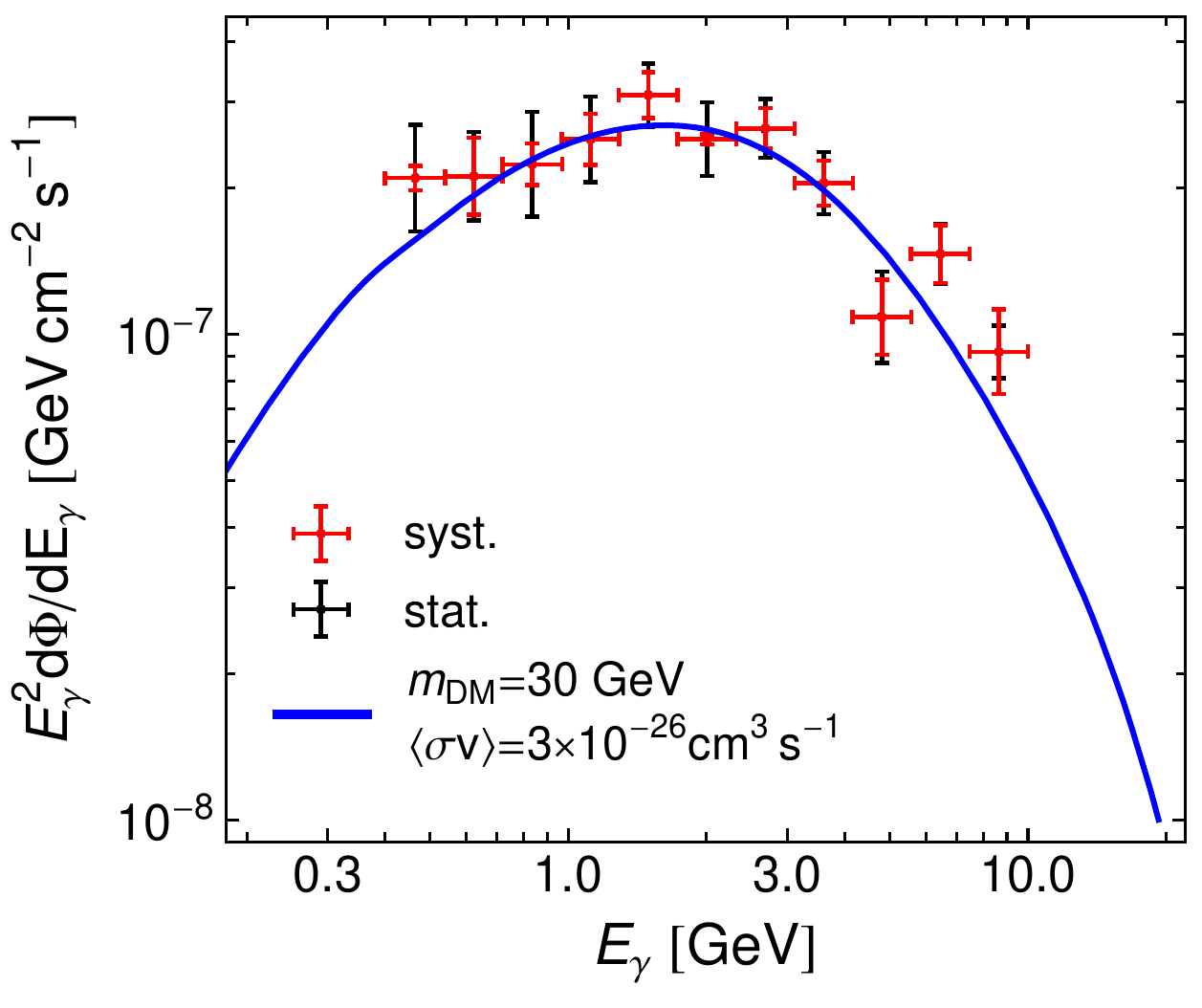}
\caption{The data points show the extended gamma-ray excess from a $7\degree\times7\degree$ region centred on the galactic centre (from~\cite{Gordon:2013vta}). The red and black error bars show the systematic and statistical uncertainties respectively. The blue solid line shows the photon spectrum corresponding to 30~GeV dark matter with an annihilation cross-section that gives the observed relic density. The branching ratios are determined by the Yukawa couplings $y_f$.
\label{fig:spectrum}}
\end{figure}

In addition to dark matter annihilation, it has been suggested that interactions between cosmic rays and gas~\cite{Linden:2012iv,Linden:2012bp,YusefZadeh:2012nh} or an unresolved population of millisecond pulsars~\cite{Abazajian:2010zy,Hooper:2011ti,Abazajian:2012pn,Mirabal:2013rba} can explain the excess. However, more detailed studies have raised problems with both of these explanations~\cite{Hooper:2013nhl,Macias:2013vya}. It is also possible that a new mechanism not proposed is responsible, since the galactic centre is a complex astrophysical environment~\cite{Cirelli:2013mqa}. For the purpose of this work, we assume that all of the excess is a result of dark matter annihilation. We use the results from the analysis of~\cite{Gordon:2013vta} (listed in their Appendix~A), who considered all events within a $7\degree\times7\degree$ region centred on the galactic centre (the position of Sgr~$\mathrm{A}^*$). Galactic backgrounds were modelled with the standard LAT diffuse model, with isotropic residuals assumed for instrumental and extragalactic sources. After background subtraction the extended emission component that they find is shown in fig.~\ref{fig:spectrum}, where the red and black error bars correspond to systematic and statistical uncertainties respectively. 

To proceed with the dark matter interpretation, it is necessary to specify the dark matter halo profile. While it is well determined far from the galactic centre, the slope is uncertain at small radii; typically there are no observations below 1~kpc and the resolution of numerical simulations is $\sim0.1$~kpc. The Einasto~\cite{Graham:2005xx} and Navarro, Frenk and White (NFW)~\cite{Navarro:1995iw} profiles are traditionally used as benchmark profiles as they provide good fits to dark matter numerical simulations~\cite{Navarro:2008kc}. However, it is possible that the dark matter halo profile remains divergent close to the centre such that profiles may behave as $\rho\propto r^{-\gamma}$ with $\gamma > 1$ ($\gamma=1$ in the NFW profile). As an example, the Via Lactea~II simulation favours a profile with $\gamma=1.24$~\cite{Diemand:2008in}. Given that the $\gamma$-ray emission traces the morphology of the profile, the consequence of a more strongly peaked profile in terms of indirect detection is a much brighter gamma-ray emission relative to the case of an Einasto or NFW profile. For the extended gamma-ray excess, it is found that a generalised NFW profile
\begin{equation}
\rho(r)=\rho_s\left(\frac{r}{r_s} \right)^{-\gamma}\left[1+\left(\frac{r}{r_s} \right) \right]^{\gamma-3}\;.
\label{eq:genNFW}
\end{equation}
with  $\gamma=1.2$ gives the best fit~\cite{Gordon:2013vta}.

The following simplified model gives a good fit to the extended gamma-ray excess shown in fig.~\ref{fig:spectrum}. We take the dark matter $\chi$ to be a Dirac fermion with mass $\mDM$ which interacts with a pseudoscalar $a$ with mass $m_a$ through the coupling $g_{\rm{DM}}$:
\begin{equation}
\mathcal{L}\supset-i \frac{g_{\rm{DM}}}{\sqrt{2}} a \bar{\chi}\gamma^5 \chi-i\sum_f\frac{g_{f}}{\sqrt{2}}a \bar{f}\gamma^5 f+\mathrm{h.c.}
\label{eq:model}
\end{equation}
The pseudoscalar couples to the Standard Model fermions with $g_f$, which we assume is equal to the Standard model Yukawa coupling \mbox{$g_f=y_f\equiv m_f/174\text{ GeV}$}. This relation is common for pseudoscalars, motivated from the minimal flavour violation (MFV) ansatz~\cite{D'Ambrosio:2002ex}.

The photon flux $\Phi$ at Earth from a region $\Delta \Omega$, assuming prompt photon emission arising from annihilation of Dirac dark matter, is~\cite{Cirelli:2010xx}
\begin{equation}
\frac{d \Phi}{d E_{\gamma}}=\frac{1}{4}\frac{r_{\odot}}{4 \pi} \left(\frac{\rho_{\odot}}{m_{\rm{DM}}}\right)^2 \langle J \rangle \Delta \Omega \sum_f \langle\sigma v\rangle_f \frac{d N_{\gamma}^f}{d E_{\gamma}}\;,
\end{equation}
where $r_{\odot}=8.25$~kpc is the distance from the galactic centre to the Earth, $\rho_{\odot}=0.42$~GeVcm$^{-3}$ is the local dark matter density~\cite{Salucci:2010qr,Iocco:2011jz}, $\langle\sigma v\rangle_f$ is the annihilation cross-section to $\bar{f}f$ and $d N_{\gamma}^f/d E_{\gamma}$ is the energy spectrum of photons produced per annihilation to $\bar{f}f$. We use the tabulated values of $d N_{\gamma}^f/d E_{\gamma}$ from~\cite{Cirelli:2010xx,Ciafaloni:2010ti}, which are generated with \texttt{PYTHIA~8.135}~\cite{Sjostrand:2007gs} and disregard any contribution to the flux that is not prompt i.e.\ we neglect all photons generated by the propagation of cosmic rays. The average $J$ factor over a region of size $\Delta \Omega$ is
\begin{equation}
\langle J \rangle=\frac{1}{\Delta \Omega}\int  \cos b \,J (b,l)\,db\, dl\;,
\end{equation}
where
\begin{equation}
J(b,l)=\int_{\rm{l.o.s}}\frac{ds}{r_{\odot}}\left.\left(\frac{\rho(r)}{\rho_{\odot}} \right)^2\right|_{r=\sqrt{r_{\odot}^2+s^2-2 r_{\odot}s \cos b\cos l}}
\end{equation}
and $s$ varies over the line of sight. We use the form of $\rho(r)$ in eq.~\eqref{eq:genNFW} with $\gamma=1.2$, $r_s=23.1$~kpc and $\rho_s$ is chosen so that $\rho(r_{\odot})=\rho_{\odot}$. Following~\cite{Gordon:2013vta}, we calculate $\langle J \rangle$ in the $7\degree\times 7\degree$ region by summing over pixels of size $0.1\degree\times 0.1\degree$. 

For the simplified model in eq.~\eqref{eq:model}, the s-wave annihilation cross-section for $\bar{\chi}\chi\to\bar{f}f$ is
\begin{align}
\langle\sigma v\rangle_f=\frac{N_C }{8 \pi}\frac{ y_f^2\, g_{\rm{DM}}^2 \mDM^2}{(m_a^2-4 \mDM^2)^2+m_a^2 \Gamma_a^2}\sqrt{1-\frac{m_f^2}{\mDM^2}}
\end{align}
where $N_C=3\,(1)$ for coloured (colour-neutral) particles and $\Gamma_a$ is the pseudoscalar width. Among the possible final states, the dominant annihilation channel is to $b$ quarks; the branching ratio to a particular final state is determined by $y_f$, for which $y_b$ is the largest.

\begin{figure}[t!]
\centering
\includegraphics[height=0.84\columnwidth]{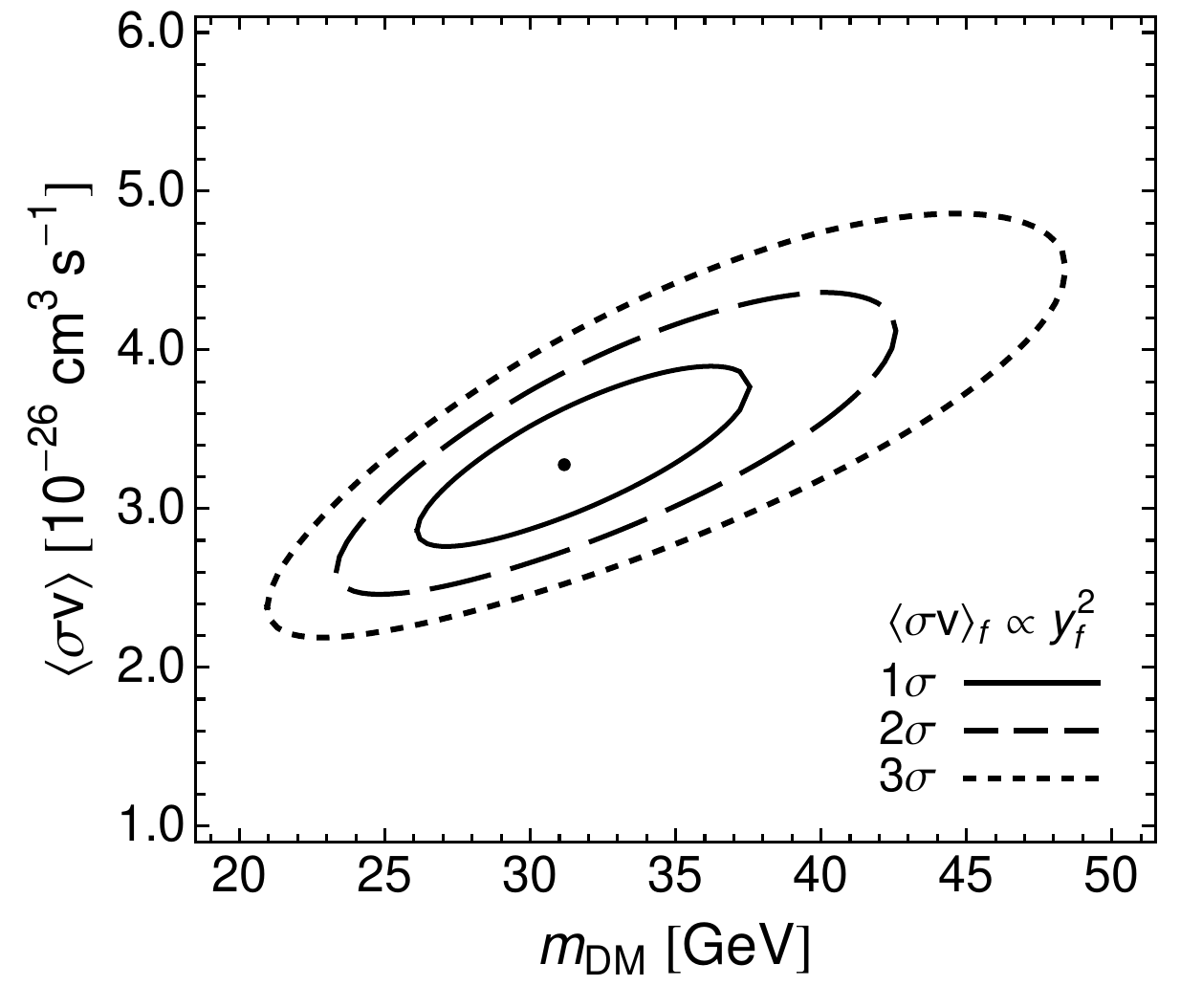}
\caption{The solid, dashed and dotted contours show the 1, 2 and 3$\sigma$ favoured regions in the $m_{\rm{DM}}$-$\langle\sigma v\rangle$ plane, along with the best fit point, shown by the dot. The branching ratios are determined by the Yukawa couplings $y_f$. The excess is consistent with an annihilation cross-section that gives the observed dark matter relic density.
\label{fig:region}}
\end{figure}

An example of the resulting gamma-ray spectrum for \mbox{$m_{\rm{DM}}=30$~GeV}, $\langle \sigma v \rangle\equiv\sum_f\langle \sigma v \rangle_f=3\times10^{-26}~\text{cm}^3\,\text{s}^{-1}$ and the astrophysical parameter choices above is shown by the solid blue curve in fig.~\ref{fig:spectrum}. This gives a good fit to the data. Being more quantitive, fig.~\ref{fig:region} shows the result of a fit in the $\mDM$ - $\langle \sigma v \rangle$ plane assuming that the branching ratio into the final state $\bar{f}f$ is determined by the Yukawa couplings $y_f$. The black dot shows the best fit point and the solid, dashed and dotted lines show the 1, 2 and 3~$\sigma$ regions respectively. These regions are determined by minimising a $\chi^2$ distribution as described in~\cite{Gordon:2013vta}. We see that the cross-section is consistent with that required for a thermal relic, i.e.\ $\langle \sigma v \rangle\simeq3\times10^{-26}~\text{cm}^3\,\text{s}^{-1}$, for $\mDM$ around 30~GeV. In addition, one should not discount the possibility that $\langle \sigma v \rangle \gg 3\times10^{-26}~\text{cm}^3\,\text{s}^{-1}$ in the primordial Universe since regeneration mechanisms, such as those proposed in \cite{Hall:2009bx,Williams:2012pz}, may maintain the would-be candidate as the main dark matter component.

\begin{figure}[t!]
\centering
\includegraphics[height=0.84\columnwidth]{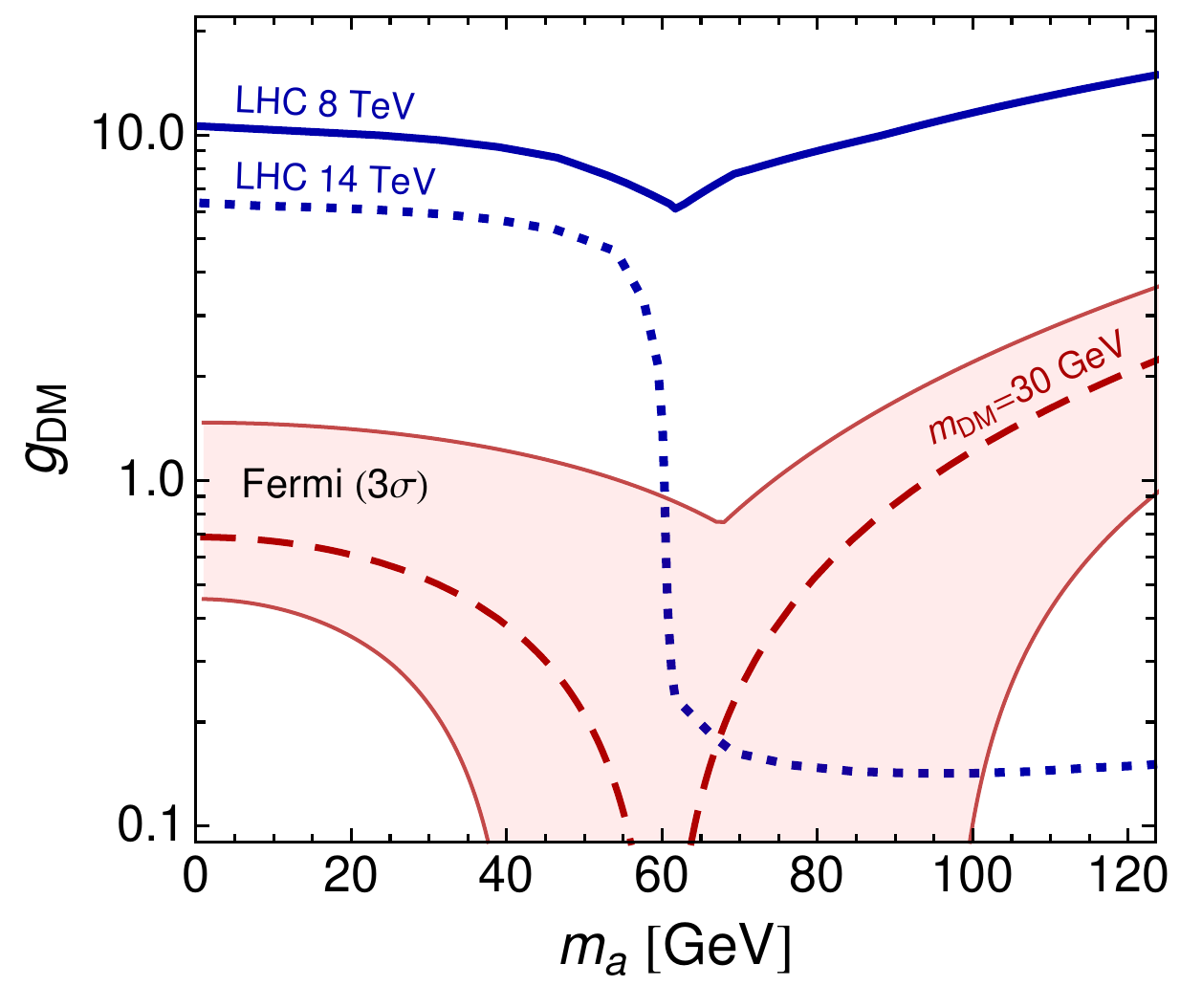}
\caption{The red shaded region shows the values of $g_{\rm{DM}}$ and $m_a$ that fit the galactic excess at $3\sigma$ (marginalising over $m_{\rm{DM}}$). The red dashed line shows the values of $g_{\rm{DM}}$ and $m_a$ that give $\langle \sigma v \rangle=3\times10^{-26}~\text{cm}^3\,\text{s}^{-1}$ for $m_{\rm{DM}}=30$~GeV. The solid blue line shows the constraint from the current 8~TeV CMS monojet search, and the blue dashed line our extrapolation of a similar search at 14~TeV with $40~\mbox{fb}^{-1}$. \label{fig:coupling}}
\end{figure}

The red shaded region in fig.~\ref{fig:coupling} shows the values of the pseudoscalar-dark matter coupling $g_{\rm{DM}}$ and mass $m_a$ that fit the galactic excess at $3\sigma$. In this region we have marginalised over $m_{\rm{DM}}$. The red dashed line shows the values of $g_{\rm{DM}}$ and $m_a$ that result in $\langle \sigma v \rangle=3\times10^{-26}~\text{cm}^3\,\text{s}^{-1}$ for $m_{\rm{DM}}=30$~GeV. Typically, a coupling of order one or less is required to fit the excess. The annihilation is resonantly enhanced when $m_a\approx2 m_{\rm{DM}}$, explaining the `funnel' that extends to small values of~$g_{\rm{DM}}$. We find that the width of the pseudoscalar varies from a few MeV to a few GeV over the parameter space. For $\mDM=30$~GeV and $(m_a,g_{\rm{DM}})=(40,0.4)$, the width is $\Gamma_a=1.9$~MeV and the largest branching ratio is $\mathrm{BR}(a\to b\bar{b})=89\%$, followed by $c\bar{c}$ and $\tau^+\tau^-$ at 7\% and 4\% respectively. Once it is kinematically possible for the pseudoscalar to decay into dark matter, this channel dominates. For instance, for the point $\mDM=30$~GeV and $(m_a,g_{\rm{DM}})=(90,1.0)$ the width is $\Gamma_a=1.3$~GeV with $\mathrm{BR}(a\to \chi\chi)=99.7\%$ and $\mathrm{BR}(a\to b\bar{b})=0.3\%$.

\section{Collider searches}
\label{sec:LHC}

In general, it is hard to find evidence for this model at a collider, particularly for a pseudoscalar that satisfies $m_a > m_h/2$ so that constraints from $h\to aa$ decays are forbidden. We have implemented our model of Dirac fermion dark matter with a pseudoscalar mediator using \texttt{FeynRules}~\cite{Christensen:2008py} with the UFO output~\cite{Degrande:2011ua} to generate events in \texttt{MadGraph5}~\cite{Alwall:2011uj}. We include the dimension five $G_{\mu\nu}\tilde{G}^{\mu\nu}a$ operator, which is obtained from integrating out the top-quark loop. To check our implementation, we compare our cross-section for $\ttbar a$ and the inclusive $pp \to a$ cross-section with those available for pseudoscalar Higgs bosons in the literature. We find good agreement with the results of the LHC Higgs Cross-Section Working Group~\cite{Heinemeyer:2013tqa}. 

We find that the greatest sensitivity comes from the 8~TeV CMS monojet search using $19.5~\textrm{fb}^{-1}$ of data~\cite{CMS-PAS-EXO-12-048}. The 90\% confidence limit we derive from that search is shown as the solid blue line in fig.~\ref{fig:coupling}. There is a constraint only at large values of the coupling $g_{\rm{DM}}$ and this search does not cut into the preferred Fermi-LAT region of good fit. The relative weakness of the LHC limit is a good demonstration of how a naive expectation of the limit based on crossing symmetry fails~\cite{Profumo:2013hqa}. It is likely that including the dimension five $G_{\mu\nu}\tilde{G}^{\mu\nu}a$ operator, rather than performing a loop calculation, overestimates the production cross-section with the result that our limit on $g_{\rm{DM}}$ is an overestimate~\cite{Haisch:2012kf}. We also note that at such large values of $g_{\rm{DM}}$, the mediator width is larger than its mass, making the particle interpretation of the mediator questionable~\cite{Buchmueller:2013dya}. It is this fact that explains the shape of the exclusion contour, since once the mediator can decay to dark matter, the mediator width increases by a factor of $\mathcal{O}(10^3)$, which suppresses the production cross-section. This limit assumes that $\mDM=30$~GeV but other values of $\mDM$ consistent with the excess will give a similar result. The magnitude of the limit will remain the same but the strongest constraint on $g_{\rm{DM}}$ will shift to $m_a\approx2 \mDM$.

We also provide a rough estimate of how monojet results at 14~TeV will affect this scenario. To do this we assume that CMS will continue using the $\MET=400$~GeV bin. As the expected backgrounds (mostly from $Z(\to \nu\nu)+1\rm{j}$) in this bin will increase, we assume that the limit on the number of monojet events will increase in such a way that S/B will remain approximately constant. The blue dashed line in fig.~\ref{fig:coupling} shows the results we obtain for an integrated luminosity of $40~\rm{fb}^{-1}$ at 14~TeV, representative of about two years running. The improvement from the 14~TeV run looks dramatic, however it is important to realise that the monojet search is not particularly sensitive to $g_{\rm{DM}}$ when the pseudoscalar is produced on-shell, as is the case when $m_{a} > 2 \mDM$. In this case the monojet plus missing energy cross-section is approximately $\sigma(pp\to a+\mathrm{j})\mathrm{BR}(a\to \chi\chi)$. For $g_{\rm{DM}}>y_b$ the branching ratio is almost 100\%, and so if a particular point in parameter space is ruled out, we would expect it to be ruled out for $g_{\rm{DM}}$ larger than the bottom Yukawa. Indeed, this is what appears. The production cross-section for the pseudoscalar plus a hard jet increases by up to a factor of seven at 14 TeV due to the large increase in the gluon PDF. For instance, for $(m_a, g_{\rm{DM}})=(100,1.0)$ we find that the monojet cross-section increases from 15~fb to 96~fb. The dominant background from $Z(\to \nu\nu)+1\rm{j}$ also increases, from 135~fb at parton level to 650~fb for $\mathrm{MET}>400$~GeV and $|\eta_{\rm{j}}|<2.4$.
We again mention that this cross-section is likely an overestimate because the top-quark loop is not taken fully into account~\cite{Haisch:2012kf}. While the monojet search is likely to start to cut into the parameter space in the $m_a\geq2\mDM$ region, the area below this is difficult to probe. 

Since the pseudoscalar mediator interacts most strongly with the top quark due to its Yukawa-like couplings, searches in the $t\bar{t}a$ final state may also be an effective means of constraining this model. A representative search is the ATLAS search for $t\bar{t}+\rm{MET}$~\cite{ATLAS-CONF-2013-048} in the dilepton final state. This search requires the $p_{T}$ of the leading lepton to be greater than 25~GeV and relies on the $\mttwo$~\cite{Lester:1999tx} variable as its main discriminant. For the main Standard Model $t\bar{t}$ background, this quantity has a kinematic edge at $m_W$. The four ATLAS search regions therefore encompass $\mttwo>90,100,110$ and $120$~GeV to suppress this. We hadronise our events using \texttt{PYTHIA~6}~\cite{Sjostrand:2006za} and pass them through the PGS~4~\cite{pgs,Cacciari:2008gp} detector simulator with an ATLAS-specific detector card, and analyse the resulting LHCO output using a modified version of Parvicursor~\cite{parvicursor}. We find that the ATLAS search has a relatively low acceptance for our model, in line with the stated ATLAS efficiencies for light top squarks in~\cite{ATLAS-CONF-2012-167}. Furthermore, the cross-sections for $t\bar{t}a$ production are known to be approximately three times smaller than for $t\bar{t}h$ production at the same mass. ATLAS set a limit in the $\mttwo=90$~GeV channel of 2.5~fb. Since this includes the leptonic top decays, this corresponds to an inclusive cross-section of 51~fb (i.e. without decaying the tops). However, the $pp\to \ttbar a$ cross-section for a 100~GeV pseudoscalar mediator is only 40~fb, so it is not surprising that that this search is not effective.
 We have cross-checked our results using the \texttt{CheckMATE}~\cite{Drees:2013wra} package which incorporates the results of~\cite{Read:2000ru,Cacciari:2011ma,deFavereau:2013fsa,Cacciari:2008gp,Lester:1999tx}. We have also used \texttt{CheckMATE} to check our scenario against the \cite{ATLAS-CONF-2013-035,ATLAS-CONF-2013-047,ATLAS-CONF-2012-104} searches at 7 and 8~TeV and find no constraint.

We next consider searches from LEP and Tevatron. Interactions between pure pseudoscalars and massive vector bosons are suppressed. Accordingly, the limit from Higgs searches at LEP and the Tevatron which rely on the vector boson fusion (VBF) and associated production modes do not constrain our model. Instead, we look to searches which are sensitive to gluon fusion at the LHC. In~\cite{Aad:2012cfr} the ATLAS Collaboration searched for neutral BSM Higgs bosons decaying to $\mu^+ \mu^-$ and $\tau^+\tau^-$ at $\sqrt{s}=7$~TeV, presenting results for $m_{a}>100$~GeV in order to avoid large backgrounds from the $Z$-boson resonance. We have checked that this does not constrain our model in this regime. For instance, ATLAS set a limit of 20~pb on $\sigma \times \mathrm{BR}(a \to \tau^+\tau^-)$ for $m_a=100$~GeV. In our simplified model with $g_{\rm{DM}}=0.05$ we obtain a cross-section of 0.45~pb, over 40 times lower than the ATLAS limit. For larger values of $g_{\rm{DM}}$ the invisible width increases, so that the branching ratios into visible final states decrease and this search loses efficiency.

Finally, $\Upsilon$ resonance decays and searches for direct production of the mediator followed by decay to $\mu^+ \mu^-$ can be used to constrain the coupling $g_f$ for pseudoscalar mediators below 10~GeV~\cite{Aaltonen:2009rq,Aubert:2009cka}. While we assumed that $g_f=y_f$, these searches are likely to constrain $g_f \lesssim y_f$ for $m_a\lesssim7~\mathrm{GeV}$ and $g_f \lesssim 0.01 y_f$ for $m_a\lesssim5~\mathrm{GeV}$. Further details can be found in~\cite{Schmidt-Hoberg:2013hba}. These searches do not {\it a priori} rule out an interpretation to the gamma-ray excess in terms of our simplified model since a decrease in $g_f$ can be compensated by increasing $g_{\rm{DM}}$. In any case, these constraints are completely avoided by considering the region $m_a>10~\mathrm{GeV}$.

While future monojet and $B$ physics searches may constrain the parameter space with $m_a\geq2 \mDM$ and $m_a\lesssim10~\mathrm{GeV}$, we conclude that in much of the parameter space, no signal will appear at collider experiments.

\section{Direct detection and other indirect searches}
\label{sec:directdetection}

The LUX experiment~\cite{Akerib:2013tjd} currently has the world leading sensitivity for spin-independent and spin-dependent dark matter-neutron interactions in the mass range that we are interested in. For experiments planning to run in the foreseeable future, LZ, which is the successor to LUX, should provide the best sensitivity, approaching the sensitivity where the irreducible background from neutrinos dominates~\cite{Malling:2011va,Billard:2013qya,Cushman:2013zza}.

The interaction between dark matter $\chi$ and a quark $q$ is described by the effective operator
\begin{equation}
\mathcal{L}=\frac{y_{q} \,g_{\rm{DM}}}{2 m_a^2}\bar{\chi}\gamma^5 \chi\, \bar{q}\gamma^5 q \;,
\end{equation}
valid because the mediator mass $m_a$ is much greater the momentum transferred in the scattering process. In order to compare theoretical predictions with experimental results, it is necessary to match the quark-level matrix element with the nucleon-level matrix element, evaluated in the non-relativistic limit. A clear discussion of this procedure is given in~\cite{Freytsis:2010ne,Cheng:2012qr,Dienes:2013xya}, with the result that
\begin{equation}
\begin{split}
&\frac{y_q\,g_{\rm{DM}}}{2 m_a^2}\bra{\chi_f}\bar{\chi}\gamma^5 \chi\ket{\chi_i} \bra{n_f}\bar{q}\gamma^5 q\ket{n_i}\\
 &\qquad\rightarrow \frac{g_{n n a} \,g_{\rm{DM}} }{2 m_a^2}\bra{\chi_f}\bar{\chi}\gamma^5 \chi\ket{\chi_i}  \bra{n_f}\bar{n}\gamma^5 n\ket{n_i}
 \end{split}\;,
\label{eq:matrixelement}
\end{equation}
where $n$ represents either a proton or neutron and 
\begin{equation}
\frac{ g_{n n a}}{m_n}=\sum_{q=u,d,s}\frac{y_q\,\Delta q }{m_q}  -\bar{m}\left(\sum_{q=u,d,s}\frac{\Delta q}{m_q}\right)\sum_{q=u,...,t} \frac{y_q}{m_q}
\end{equation}
with $\bar{m}=(1/m_u+1/m_d+1/m_s)^{-1}$~\cite{Cheng:2012qr}. Since we are considering scattering at LZ, which has a xenon target nucleus, we ignore contributions from proton scattering because the spin of a xenon nucleus is dominantly carried by the neutron. In this case, using
\begin{equation}
\Delta u=-0.44\,,\quad \Delta d=0.84\,, \quad \Delta s=-0.03
\end{equation}
and $y_q=m_q/174$~GeV, we obtain $g_{n n a}\approx2.8\times10^{-3}$.

The non-relativistic limit of eq.~\eqref{eq:matrixelement} leads to a spin-dependent interaction; for dark matter with speed $v$, we find that the differential scattering cross-section to scatter of a nucleus of mass $m_{\rm{N}}$ with spin $J_{\rm{N}}$ and spin structure function $S_{A}(q)$~\cite{Klos:2013rwa} is
\begin{equation}
\label{eq:DDdRdE}
\frac{d \sigma}{d E_{\rm{R}}}=\frac{q^4}{\mDM^2 m_{\rm{N}}^2} \frac{3 g_{nna}^2\, g_{\rm{DM}}^2  m_{\rm{N}}}{8 m_a^4 v^2}\frac{1}{2 J_{\rm{N}}+1}S_{A}(q)\;,
\end{equation}
where $q^2=2 m_{\rm{N}}E_{\rm{R}}$ is the momentum transfer and $E_{\rm{R}}$ is the nuclear recoil energy. The typical recoil energy  under investigation at direct detection experiments is $E_{\rm{R}}\sim10~\text{keV}$ so that $q\sim100~\text{MeV}$. Crucially, we see that the factor $q^4/\mDM^2 m_{\rm{N}}^2$ suppresses the cross-section by a factor $\mathcal{O}(10^{-12})$. Owing to this, the number of expected events at LZ between 2~PE and 30~PE in the vicinity of $\mDM=30$~GeV is
\begin{equation}
N_{\rm{s}}\approx1\text{ event} \left(\frac{g_{\rm{DM}}}{1} \right)^2 \left( \frac{250\text{ MeV}}{m_a} \right)^4 \left(\frac{\mathrm{Exp}}{10^7 \text{ kg-days}} \right)\;.
\end{equation}
Here we have followed the standard procedure to calculate the number of events~\cite{Lewin:1995rx,McCabe:2013kea} and assumed that efficiencies at LZ are the same as those at LUX. 

In addition to the result above, which takes into account all of the momentum dependence in the scattering process, we also provide a reference cross-section $\tilde{\sigma}_{n,p}^{\rm{SD}}$ that can be compared directly with experimental limits. Mapping eq.~\eqref{eq:DDdRdE} onto the form that is constrained by experiments (see e.g.~\cite{Kopp:2009qt} for details), we find that
\begin{align}
\label{eq:dmnxsec}
\tilde{\sigma}_n^{\rm{SD}}&=\frac{9}{16 \pi} \frac{ q^4}{m_{\rm{DM}}^2 m_{\rm{N}}^2} \frac{g_{nna}^2\, g_{\rm{DM}}^2\,\mu_{n}^2}{m_a^4}\\
&\approx8\times 10^{-43}\text{ cm}^2 \left(\frac{g_{\rm{DM}}}{1} \right)^2 \left( \frac{250\text{ MeV}}{m_a} \right)^4\;,
\end{align}
where $\mu_n$ is the dark matter-neutron reduced mass and we have assumed that $m_{\rm{DM}}=30$~GeV and $q=100$~MeV. This cross-section is similar to the projected LZ limit \mbox{$\tilde{\sigma}_n^{\rm{SD}}\leq7\times 10^{-43}\text{ cm}^2$} at $m_{\rm{DM}}=30$~GeV that is presented in~\cite{Cushman:2013zza}, validating the result of the more precise analysis above. As mentioned previously, the spin of a xenon nucleus is dominantly carried by the neutron so we ignored the contribution from the proton spin. In contrast, the proposed PICO250 experiment (a joint experiment from the COUPP and PICASSO collaborations) is more sensitive to the dark matter-proton scattering cross-section $\tilde{\sigma}_p^{\rm{SD}}$ and they estimate that they will exclude scattering cross-sections smaller than $8\times10^{-43}$~cm$^2$ at $m_{\rm{DM}}=30$~GeV~\cite{Cushman:2013zza}. The dark matter-proton cross-section takes the same form as eq.~\eqref{eq:dmnxsec} except \mbox{$g_{ppa}\approx-1.1\times10^{-2}$} should be used instead of $g_{nna}$. With $q=50$~MeV, appropriate for scattering off flourine, we find that
\begin{equation}
\tilde{\sigma}_p^{\rm{SD}}\approx8\times 10^{-43}\text{ cm}^2 \left(\frac{g_{\rm{DM}}}{1} \right)^2 \left( \frac{650\text{ MeV}}{m_a} \right)^4\;,
\end{equation}
so that PICO250 will set a slightly stronger constraint on $m_a$ than LZ.

Even with the large exposure collected by LZ and PICO250 (a factor $10^3$ larger than the current exposure of LUX), we find that LZ and PICO250 would only begin to observe events for $m_a\lesssim250$~MeV and $m_a\lesssim650$~MeV respectively. For heavier values of the pseudoscalar mass, the number of events drops rapidly so that there is no possibility of LZ or PICO250 observing any events from dark matter scattering for $m_a\gtrsim10$~GeV. As the tree level contribution is strongly suppressed, we should consider if the loop-induced spin-independent interaction gives a larger contribution. The authors of~\cite{Freytsis:2010ne} considered this possibility and found that the spin-independent interaction is smaller than the tree level contribution considered above. Therefore, we conclude that direct detection experiments cannot probe this scenario.

Finally, we consider other indirect searches for WIMP dark matter. Firstly, limits from the anti-proton flux (derived from low-energy data collected by BESS-Polar~II~\cite{Abe:2011nx}) exclude a thermal WIMP that dominantly annihilates to quarks when $\mDM=3\text{-}20$~GeV~\cite{Kappl:2011jw}, which is below the mass range favoured by the gamma-ray excess in fig.~\ref{fig:region}. Using the anti-proton flux calculated with~\cite{Cirelli:2010xx}, we also checked that the limit derived from the anti-proton flux at higher energy, as measured by PAMELA~\cite{Adriani:2010rc}, does not exclude the favoured region, in agreement with~\cite{Kappl:2011jw, Cirelli:2013hv}. Secondly, the cosmic microwave background (CMB) provides constraints from the energy deposition arising from dark matter annihilation~\cite{Chen:2003gz}. However, these constraints are weakened when the dominant annihilation channel is to heavy quarks or $\tau$ leptons, with the result that current and projected limits do not constrain this model~\cite{Cline:2013fm,Madhavacheril:2013cna}. Thirdly, limits from the neutrino flux from dark matter annihilation in the Sun are not applicable because the capture cross-section from scattering on protons, 
\begin{equation}
\sigma_{\rm{SD}}^p\approx2\times 10^{-43}~\text{cm}^2 \left(\frac{g_{\rm{DM}}}{1}\right)^2 \left(\frac{1~\text{GeV}}{m_a}\right)^4\;,
\end{equation}
is orders of magnitude below the limit of $10^{-38}\text{ cm}^2$ from Super-Kamiokande~\cite{Desai:2004pq}. Here we assumed that \mbox{$\mDM=30$~GeV} and \mbox{$q=20$~MeV}, typical for a scattering event in the Sun. Fourthly, Fermi-LAT limits from dwarf spheroidal galaxies are unlikely to definitively detect or reject the dark matter hypothesis~\cite{Macias:2013vya}. Therefore, we conclude that additional indirect detection signatures do not provide further constraints.

\section{Conclusions}
\label{sec:conclusions}

If dark matter is a WIMP with weak-scale interactions with the Standard Model particles, then the prospects of discovery at direct detection, indirect detection or collider experiments are good. In many models of WIMP dark matter, if a signal is produced in one experimental channel, then a signal will also be observed in another. However, we show that this need not be the case and that dark matter may be coy, producing a single large observable signal in isolation.

We demonstrate this by considering the extended gamma-ray excess from the galactic centre, observed by the Fermi-LAT satellite. Although the origin of this excess is uncertain, one way to account for it is with WIMP dark matter annihilating dominantly to $b$ quarks in the galactic centre. We showed that a simple model of Dirac dark matter that is coupled to the Standard Model through a pseudoscalar, which has couplings to the Standard Model particles that are proportional to the Yukawa couplings, can account for the excess (see fig.~\ref{fig:spectrum}). A fit to the excess shows that the preferred dark matter mass $\mDM$ is between $\sim20$-50~GeV and that the annihilation cross-section is consistent with $\langle \sigma v\rangle\simeq3\times10^{-26}~\text{cm}^3\,\text{s}^{-1}$ (see fig.~\ref{fig:region}), required for the dark matter to obtain its relic abundance through the thermal freeze-out mechanism. This cross-section implies that the dark matter-pseudoscalar coupling $g_{\rm{DM}}$ is $\mathcal{O}(1)$ or less over a large range of pseudoscalar mass $m_a$ (see fig.~\ref{fig:coupling}).

Finding additional experimental evidence for this simple model is difficult. From colliders, the greatest sensitivity comes from the CMS monojet search. Although this search does not currently constrain any of the favoured parameter space, the projected limit from the 14~TeV LHC run constrains the region $m_{a}\gtrsim2\mDM$ (see fig.~\ref{fig:coupling}). Owing to the suppressed dark matter-nucleus interaction, future direct detection experiments have no sensitivity when $m_{a}\gtrsim190$~MeV. Furthermore, additional indirect searches in the anti-proton flux, the CMB, the neutrino flux from the Sun and the photon flux from dwarf spheroidal galaxies do not provide further constraints.

Therefore, over much of the parameter space, the extended gamma-ray excess exists in isolation as the sole evidence for particle dark matter. For WIMPs that produce observable signals in isolation, our results emphasise the importance of fully understanding that signal. In the case of the extended gamma-ray excess, it is crucial that additional hypothesises with an astrophysical origin are fully explored so that they may excluded. 
 
\section*{Acknowledgements}

MJD thanks Randel Cotta and Alex Wijangco for comparing results with him.  CM thanks Alastair Currie, Jonathan Davis and Felix Kahlhoefer for discussions regarding LUX, Takashi Toma for discussions on the velocity dependence of annihilation cross-sections and Kai Schmidt-Hoberg for discussions on constraints from $B$ searches. CJW thanks ITP Heidelberg for hospitality while some of this work was carried out. This work has been partially supported by the European Union FP7 ITN INVISIBLES (Marie Curie Actions, PITN- GA-2011- 289442).


\bibliography{ref}
\bibliographystyle{arxiv}

\end{document}